\documentclass{article}
\usepackage{spconf,amsmath,graphicx}
\usepackage{booktabs}
\usepackage{multirow}
\usepackage{multicol}
\usepackage{hyperref}
\usepackage{xcolor}
\usepackage{color}
\usepackage{verbatim}
\hypersetup{
    colorlinks=true,
    linkcolor=blue,
    filecolor=magenta,      
    urlcolor=blue,
}

\newcommand{\be}[1]{\textbf{#1}}
\newcommand{\se}[1]{\underline{#1}}

\usepackage{soul}
\usepackage{color}

\makeatletter
\renewcommand\section{\@startsection{section}{1}%
     {0pt}
     {-\baselineskip}
     {2.0ex}
     {}}
\renewcommand\subsection{\@startsection{subsection}{2}%
     {0pt}
     {-1.5ex}
     {1.6ex}
     {}}
\renewcommand\subsubsection{\@startsection{subsubsection}{3}%
     {0pt}
     {-1ex}
     {.5ex}
     {}}
\makeatother

\title{StreamVC: Real-Time Low-Latency Voice Conversion} 
\name{Yang Yang, Yury Kartynnik, Yunpeng Li, Jiuqiang Tang, Xing Li, George Sung, Matthias Grundmann}
\address{Google LLC, U.S.A.}
\begin{document}
%
\maketitle
\begin{abstract}
We present StreamVC, a streaming voice conversion solution that preserves the content and prosody of any source speech while matching the voice timbre from any target speech. Unlike previous approaches, StreamVC produces the resulting waveform at low latency from the input signal even on a mobile platform, making it applicable to real-time communication scenarios like calls and video conferencing, and addressing use cases such as voice anonymization in these scenarios. Our design leverages the architecture and training strategy of the SoundStream neural audio codec for lightweight high-quality speech synthesis. We demonstrate the feasibility of learning soft speech units causally, as well as the effectiveness of supplying whitened fundamental frequency information to improve pitch stability without leaking the source timbre information.

\end{abstract}
\section{Introduction}
\label{sec:intro}

\emph{Voice conversion} refers to altering the style of a speech signal while preserving its linguistic content. While style encompasses many aspects of speech, such as emotion, prosody, accent, and whispering, in this work we focus on the conversion of speaker timbre only while keeping the linguistic and para-linguistic information unchanged.

Early attempts at voice conversion rely on the idea of CycleGAN- or StarGAN-based \cite{starganv2} direct conversion, or auto-encoding with learned feature disentanglement. Both, however, fail to deliver high-quality results. The former empirically suffers from noticeable artifacts, and the latter mostly relies on creating information bottlenecks, either at the latent \cite{autovc, VQMIVC} or architecture level \cite{AdaINVC}, which are difficult to tune: having such a bottleneck too wide leads to leakage of source speaker information, while making it too narrow degrades content fidelity.

Recent solutions \cite{SVC_IBF, ppg_vc, SoftVC, FreeVC, QuickVC, a2a} converge to a design where the content information is obtained by leveraging pre-trained feature extraction networks either from speech recognition systems, referred to as phoneme-posteriorgram (PPG) approach \cite{SVC_IBF, ppg_vc, a2a}, or from self-supervised representation learning. Specifically, \cite{SoftVC, QuickVC} leverage HuBERT \cite{hubert}, and \cite{FreeVC} utilizes WavLM \cite{wavlm}. The combination of content information and a learned global speaker embedding serves as input and conditioning to some vocoder models, such as that used in \cite{vits, hifigan, melgan},  which are trained to reconstruct the audio waveform.

Our proposal follows the same design pattern as \cite{SoftVC, QuickVC} and uses the pseudo-labels derived from HuBERT \cite{hubert} to learn a content encoder that outputs soft speech units. The contributions and new design elements of our solution are as follows:
\begin{enumerate}
\setlength\itemsep{0pt}
    \item[(1)] We demonstrate the feasibility of using a lightweight causal convolutional network to capture the soft speech unit information instead of the existing approach \cite{SoftVC,QuickVC} of using a computationally demanding non-causal multi-layer transformer network.
    \item[(2)] By leveraging the architecture and training strategy of the SoundStream neural audio codec \cite{soundstream}, we achieve high quality speech synthesis while explicitly addressing the problem of on-device low-latency streaming inference, which was not addressed in previous work in this area.\footnotemark
    \item[(3)] We introduce the injection of whitened $f_0$ (fundamental frequency) information, which improves pitch consistency without leaking the source speaker timbre.
\end{enumerate}

\footnotetext{Note that \cite{SVC_IBF} also tackles the streaming VC problem, but it is a solution trained for a single target timbre with an inference latency of 270\,ms on a desktop CPU, whereas we achieve 70.8\,ms latency on a mobile device.}

As a result, we achieve a low inference latency of 70.8\,ms relative to the input signal on the Pixel 7 smartphone while obtaining conversion quality on par with or better than the existing state of the art\footnotemark.

\footnotetext{Audio samples available at \href{https://google-research.github.io/seanet/stream_vc/}{google-research.github.io/seanet/stream\_vc/}}

\begin{figure*}[!t]
\centering
\includegraphics[width=0.8\textwidth]{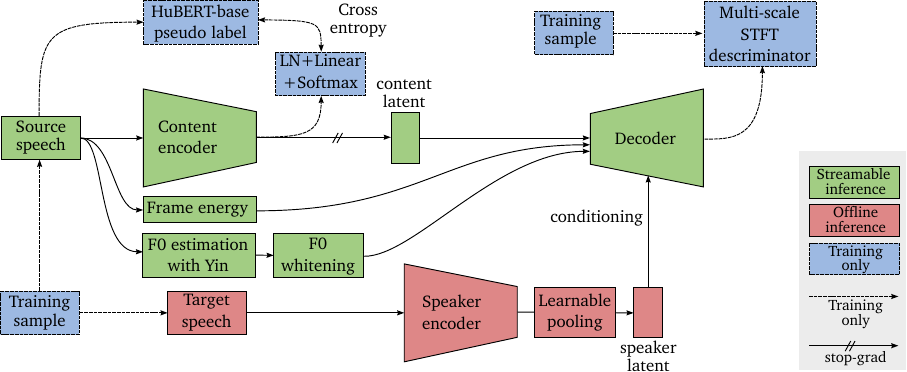}
\caption{System diagram of StreamVC.}
\label{fig:network}
\end{figure*}

\vspace{-0.0cm}
\section{Method}
\label{sec:method}

\subsection{Motivation}

Our design is inspired by Soft-VC\cite{SoftVC} and SoundStream\cite{soundstream}. 

We follow Soft-VC \cite{SoftVC} in deriving \emph{soft speech units} by using \emph{discrete speech units} extracted from HuBERT \cite{hubert} as the prediction target for a content encoder network.
Specifically, we use a pretrained HuBERT model to derive target per-frame pseudo-labels (in the same manner as \cite{SoftVC}) for the training of the content encoder, whose last layer activation before the layer norm and the logistic projection is considered as the latent soft speech units representing the speech content and passed to the decoder. 
The architecture of the content encoder and the decoder, as well the training strategy, adopt that of the SoundStream neural audio codec \cite{soundstream} for high quality causal audio synthesis.

\subsection{Model Architecture}

The overall network architecture is outlined in Fig.~\ref{fig:network}, which we detail next.

\subsubsection{Content Encoder}

The content encoder model architecture is convolutional. Its design follows the SoundStream encoder with the scale parameter $C = 64$ and the embedding dimensionality $D = 64$ (please refer to Fig.~3 in \cite{soundstream} for the details), except that there is no Feature-wise Linear Modulation (FiLM) \cite{film} layer since no conditioning needs to be applied.

\subsubsection{Speaker Encoder}

The speaker encoder consists of a per-frame encoding part and a global (utterance-level) context aggregator. The former reuses the SoundStream encoder design with the scale parameter $C = 32$ and the same embedding dimensionality $D = 64$ (please refer again to \cite{soundstream}, Fig.~3).

These per-frame context embeddings are subsequently aggregated into a single global context via \emph{learnable pooling}, which is a weighted generalization of average pooling whose weights are derived from an attention mechanism with a single learnable query. 

\subsubsection{Fundamental Frequency and Energy Estimation}

We notice that when the decoder is provided with the HuBERT-derived soft speech unit embeddings alone, the produced sound tends to have a flattened pitch envelope, having a detrimental effect on speech intonation (see section~\ref{sec:ablation}). Since the tonal and acoustic energy information are unlikely to be present in phonetic unit discrimination, we supply the decoder with these signals separately alongside the content encoder output.

For pitch (known as fundamental frequency $f_0$) estimation, we adopt the Yin algorithm \cite{yin}. Besides the $f_0$ estimate, we also supply two parameters coming from this algorithm: the cumulative mean normalized difference value at the estimated period and the estimated unvoiced (aperiodic) signal predicate. In order to convey uncertainty information regarding the $f_0$ estimation, we concatenate the outputs of the Yin algorithm with 3 different thresholds of the difference function: 0.05, 0.1, and 0.15. This produces a total of 9 values per 20\,ms concatenated in channel dimension with the content latent.

To avoid suggesting speaker timbre parameters to the decoder through this channel, we normalize the $f_0$ envelopes based on utterance-level mean and standard deviation during training and evaluation. During streaming inference, to maintain causality, running averages of these parameters are employed instead. Aside from the $f_0$ information, we also supply the energy of each 20\,ms audio frame, measured via sample variance, as side information to the decoder.

\subsubsection{Decoder}

We follow the design of the SoundStream decoder (right-hand side of Fig.~3 in \cite{soundstream}) with $C=40$ as the scale and $D=64$ as the embedding dimensionality.
Feature-wise Linear Modulation (FiLM) layers \cite{film} are likewise used in between residual units to integrate the conditioning signal in the form of the speaker latent embedding. They transform neural network features with scale and bias parameters computed from two separate linear layers taking the speaker latent as input.

\subsection{Training Strategy}

\subsubsection{Soft-Label Creation and Content Encoder Training}

To produce the soft speech units for content encoder training, we follow the procedure of Soft-VC \cite{SoftVC}. We extract the 7th transformer layer activation from a pre-trained HuBERT-Base model \cite{hubert}. (We confirmed the observation that these features yield good performance on phone discrimination tests.) Next we apply mini-batch $k$-means clustering to find the 100 centroids that are subsequently used for nearest-centroid vector quantization, defining 100-class pseudo-labels as learning targets for the content encoder at the frequency of 50 Hz.

We prevent the gradient flow from the decoder back into the content encoder. This has proven to be essential to make sure the content encoder doesn't learn to leak additional speaker information through the content latent embeddings, bypassing the speaker latent.

\subsubsection{Training Loss}

We employ a combination of adversarial (GAN) loss, feature loss, and reconstruction loss (all following \cite{melgan, li_seanet_bwe_icassp_2021}), as well as a cross-entropy loss for the content encoder latent projection in the HuBERT pseudo-label prediction. The weighting on the first three losses is inherited from \cite{soundstream}.

\subsection{Real-Time Inference}

\begin{figure}[!t]
\centering
\includegraphics[width=0.40\textwidth]{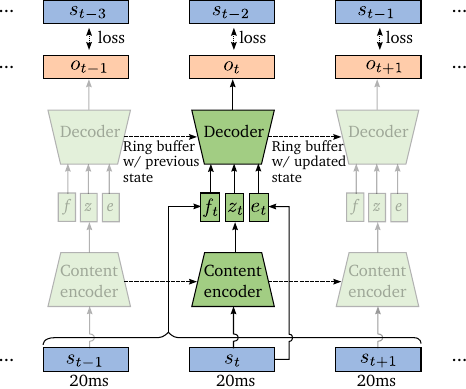}
\caption{Streaming inference.}
\label{fig:streaming}
\end{figure}

\subsubsection{Streaming Inference}

To enable online inference, we leverage the streaming-aware convolution modules introduced in \cite{kws_streamable} and extended in \cite{li_seanet_bwe_icassp_2021}.
All the convolution layers in our model are causal, but a certain architectural latency comes from the presence of strided and transposed convolutions with a limited lookahead. 
Given that the temporal resolution of the content latent is 320 times coarser than the audio, the minimum inference frame size is 320 samples, which is 20\,ms of audio sampled at 16\,kHz. The network inference is correspondingly triggered at the rate of 50 Hz. 

\subsubsection{Lookahead and Architectural Latency}

As shown in Fig.~\ref{fig:streaming}, a 2-frame lookahead is introduced by pairing the output frame $o_t$ with the input frame $s_{t-2}$ for loss computation. The $f_0$ information $f_t$ passed into the decoder at time step $t$ is computed with a context window (maintained as a ring buffer during inference) that spans three frames $(s_{t-1}, s_{t}, s_{t+1})$. This translates into a 60\,ms architectural latency since the inputs up to the time step $t+1$ are required to compute the output for the time step $t-2$.

\subsubsection{Computational Latency}
To profile inference latency, we use XNNPACK \cite{xnnpack}, a highly optimized CPU inference backend. On average, running the content encoder and the decoder on a single CPU core of a Pixel 7 smartphone\footnote{We observed similar performance on iPhone 8.} takes 10.8\,ms\footnote{In which the Yin algorithm takes negligible sub-millisecond time, since the most computationally heavy operations for it are FFT and IFFT for the auto-correlation computation.} for each 20\,ms chunk of audio. It is tested that the entire pipeline can run continuously in real time in a streaming fashion. The end-to-end latency, a combination of architectural and inference latency, is thus 70.8\,ms in this environment.

\begin{table*}[t]
\centering
\begin{tabular}{ @{}l l c c c c c c c @{}} 
\toprule
 & \multirow{2}{9em}{Metrics} & \multicolumn{3}{c}{Naturalness (DNSMOS)} & \multicolumn{2}{c}{Intelligibility} & Speaker similarity & $f_0$ consistency \\
 \cmidrule(lr){3-5} \cmidrule(lr){6-7}  \cmidrule(lr){8-8} \cmidrule(lr){9-9}
& & SIG & BAK & OVRL & WER & CER & Resemblyzer score & $f_0$ PCC \\
\midrule 
\multirow{4}{4em}{Baselines (VCTK)}& VQMIVC \cite{VQMIVC} & 3.36 &	3.29 &	3.86 &	57.88\% &	35.50\% &	65.72\% &	0.680\\
& BNE-PPG-VC \cite{ppg_vc} & 3.54 &	3.42 &	3.89 &	11.43\% &	5.24\% &	80.17\% &	0.723\\
& Diff-VCTK \cite{Diff} & \se{3.57} &	\be{3.60} &	\be{4.11} &	11.64\% &	4.38\% &	\be{84.23}\% &	0.219\\
& QuickVC \cite{QuickVC} & \be{3.61} &	\se{3.59} &	\se{4.08} &	\se{6.30}\% &	3.03\% &	78.51\% &	\se{0.758}\\
\midrule 
\multirow{2}{4em}{Ours}&StreamVC (LibriTTS) & \se{3.57} & 	3.53 &	4.07 &	\be{6.22}\% &	\be{2.17}\% &	77.81\% &	\be{0.842}\\
&$+$ fine-tuning on VCTK& 3.56&	3.48&	4.02&	6.54\%&	\se{2.25}\%&	\se{80.34}\%&	0.727\\
\midrule
\midrule
\multirow{2}{4em}{Ablations}& $-$ $f_0$ whitening & 3.48	& 3.40 &	4.00 &	6.12\% &	2.04\% &	75.59\%& 	0.704\\
& $-$ $f_0$ &3.56&	3.55&	4.06&	6.97\%&	2.56\%&	77.46\%&	0.461\\

\midrule
\multirow{2}{4em}{Oracles}&Source & 3.57 &	3.56 &	3.99 &	5.41\% &	2.43\% &	53.86\% &	1.000\\
&Target & - &	- &	- &	- &	- &	82.42\% &	-\\
\bottomrule
\end{tabular}
\caption{\label{table:result}%
Evaluation results of StreamVC in comparison with the baseline approaches. 
}

\end{table*}

\section{Experiment}
\subsection{Training and Evaluation Datasets}

We use the LibriTTS \cite{libritts} \texttt{train-clean-100} subset to derive the cluster centroids for the HuBERT pseudo-labels. All LibriTTS \texttt{train} subsets form our training dataset, which contains 555.15 hours of speech from 2311 speakers. For all our experiments, a sampling rate of 16\,kHz is used. The network is trained for 1.3 million steps with a batch size of 128.

To evaluate the generalization performance of the model, we form the evaluation dataset where both the source speakers and target speakers are unseen during training. Specifically, we collect the first 10 utterances from each speaker\footnote{For a few speakers with less than 10 utterances in total, we use all the available ones.} in \texttt{test-clean} subset of LibriTTS as the source speech, which gives us 377 source utterances. We then randomly select 6 speakers\footnote{Speakers IDs are \texttt{p231}, \texttt{p334}, \texttt{p345}, \texttt{p360}, \texttt{p361}, and \texttt{p362}.} (3 male and 3 female) from the VCTK dataset and use the concatenation of first three utterances per speaker as the target speech sample. All evaluations on done on $377\times 6=2262$ pairs of source and target speech utterances.

\subsection{Baselines and Evaluation Metrics}

Four baseline models are selected to compared with StreamVC: Diff-VCTK\footnote{\href{https://github.com/huawei-noah/Speech-Backbones/tree/main/DiffVC}{https://github.com/huawei-noah/Speech-Backbones/tree/main/DiffVC}} \cite{Diff}, BNE-PPG-VC\footnote{\href{https://github.com/liusongxiang/ppg-vc}{https://github.com/liusongxiang/ppg-vc}} \cite{ppg_vc}, VQMIVC\footnote{\href{https://github.com/Wendison/VQMIVC}{https://github.com/Wendison/VQMIVC}} \cite{VQMIVC}, and QuickVC\footnote{\href{https://github.com/quickvc/QuickVC-VoiceConversion}{https://github.com/quickvc/QuickVC-VoiceConversion}} \cite{QuickVC}. All these solutions are trained on the VCTK dataset, so all 6 target speakers are  seen during training.

The evaluations are performed along four axes: naturalness, intelligibility, speaker similarity, and $f_0$ consistency. Naturalness is rated by DNSMOS \cite{dnsmos} which consists of three scores for the quality of speech (SIG), noise (BAK), and overall (OVRL). Intelligibility is assessed by word and character error rate obtained using the HuBERT-Large ASR model.\footnote{\href{https://huggingface.co/facebook/hubert-large-ls960-ft}{https://huggingface.co/facebook/hubert-large-ls960-ft}} We follow \cite{QuickVC} by using Resemblyzer\footnote{\href{https://github.com/resemble-ai/Resemblyzer}{https://github.com/resemble-ai/Resemblyzer}} to rate speaker similarity. Lastly, we evaluate $f_0$ consistency, which is essential for tone languages, via the Pearson correlation coefficient (PCC) between the $f_0$ contours of source and converted speech. 
\vspace{0.1cm}
\subsection{Results and Observations}
The evaluation results are presented in Table \ref{table:result}. Since all the baseline models are trained on VCTK, we report the performance of both our original model \emph{StreamVC (LibriTTS)} and the result of this model finetuned on the 110 VCTK speakers.

The DNSMOS score shows that Diff-VCTK, QuickVC, and StreamVC achieve similar rating, all with the overall (OVRL) results better than the source speech. Intelligibility-wise, our base StreamVC model achieves the best WER and CER, despite being the only one trained exclusively on the speakers unseen in the evaluation. The $f_0$ PCC of the original StreamVC model is also the highest at 0.842, better than the second best by almost 10\%. This indicates that StreamVC can preserve the source tone well without much hallucination thanks to the explicit conditioning with the whitened $f_0$ signal. Speaker similarity-wise, StreamVC is worse than BNE-PPG-VC and Diff-VCTK, and is similar to QuickVC (with difference less than $1\%$). After fine-tuning on VCTK, StreamVC achieves the second best score of $80.34\%$. We notice the general trend of a trade-off between speaker similarity and $f_0$ consistency: finetuning on VCTK leads to better speaker similarity but worse $f_0$ consistency; DiffVC is an extreme case with the best speaker similarity but the worst $f_0$ PCC. We attribute this trade-off to the hypothesis that a better similarity score can be achieved by the decoder model overfitting phonetic information with its common pitch contour for the small amount of VCTK speakers.

\vspace{0.1cm}
\subsubsection{Ablation Studies}
\vspace{0.2cm}
\label{sec:ablation}
To verify the effectiveness of $f_0$ injection and $f_0$ whitening, we evaluate two ablated designs on top of the base model trained on LibriTTS (see the ``Ablations" row in Table~\ref{table:result}): 

(1) Without $f_0$ whitening. In this case we remove the whitening operation and instead follow the design in \cite{a2a} (Section 3.6 thereof) by applying a random offset and scaling factor on top of the $f_0$ estimate to avoid leaking source speaker information to the decoder. Both speaker similarity and $f_0$ PCC become noticeably worse.

(2) Without $f_0$ altogether. In this case the $f_0$ PCC falls all the way down to 0.461, compared to the best case of 0.842. Subjective listening of the converted speech reveals that the output tends to suffer from a flattened pitch envelope, which suggests that the soft speech units in \cite{SoftVC} do not carry pitch contour information and verifies the necessity of injecting it.

\vspace{0.2cm}
\section{Conclusion}

We have demonstrated that voice conversion can be efficiently performed in a streaming fashion at low latency on a mobile device without significantly sacrificing the output quality. In particular, it has been shown that HuBERT-derived soft speech units are learnable by a streamable causal convolutional neural network architecture, and injecting whitened $f_0$ information to the decoder is essential in delivering high quality output.

\vspace{-0.3cm}
\section{Acknowledgements}
\vspace{-0.2cm}
\label{sec:ack}
The authors would like to thank Marco Tagliasacchi, Qiong Hu, and Lev Finkelstein for the helpful discussions and feedback on this work.

\newpage

\let\oldbibliography\thebibliography
\renewcommand{\thebibliography}[1]{\oldbibliography{#1}
\setlength{\itemsep}{0pt}} 

\bibliographystyle{IEEEbib}
\bibliography{refs, vc_refs}

\end{document}